# Atomistic simulation of the electronic states of adatoms in monolayer MoS$_2$

Jiwon Chang, Stefano Larentis, Emanuel Tutuc, Leonard F. Register and Sanjay K. Banerjee

Microelectronics Research Center, The University of Texas at Austin, Austin, TX 78758, USA

Using an *ab initio* density functional theory (DFT) based electronic structure method, we study the effects of adatoms on the electronic properties of monolayers of the transition metal dichalcogenide (TMD) Molybdenum-disulfide (MoS$_2$). We consider the 1$^{st}$ (Li, Na, K) and 7$^{th}$ (F, Cl, Br) column atoms and metals (Sc, Ti, Ta, Mo, Pd, Pt, Ag, Au). Three high symmetry sites for the adatom on the surface of monolayer MoS$_2$ are examined as starting points to search for the most energetically stable configuration for each adatom-monolayer MoS$_2$ system, as well as the type of associated bonding. For the most stable adatom positions, we characterize the emergence of adatom-induced electronic states including any dopant states.

Transition metal dichalcogenides (TMDs) have received an increased interest as a new family of two-dimensional (2D) layer materials, following graphene [1]. Bulk TMDs with the structural formula $MX_2$ (M: transition metal, X: chalcogen) are formed from multiple $MX_2$ monolayer held together by Van-der-Waals forces. Each $MX_2$ monolayer, in turn, consists of two outer sheets of chalcogen (X) atoms much more strongly bonded to an inner sheet of M atoms, with a hexagonal in-plane primitive lattice, as shown in Fig. 1(a). Similar to graphite that is composed of graphene layers, the weak Van-der-Waals forces allow exfoliation of $MX_2$ monolayers from the bulk [2,3]. Due to their atomic scale thickness, monolayers of TMDs offer a high degree of electrostatic control. Moreover, unlike graphene, monlolayer dichalcogenides have large energy band gaps. N-type field-effect-transistors (n-FETs) have been realized with high ON-OFF current ratios and good subthreshold slope in both monolayer [1] and multilayer $MoS_2$ [4,5]. Excellent subthreshold characteristics are also observed in the multilayer $MoSe_2$ [6]. Integrated circuit based on bilayer [7] and monolayer $MoS_2$ [8] have been demonstrated as well. In addition, high performance p-type FETs (p-FETs) based on the monolayer $WSe_2$ [9] have been demonstrated, suggesting the possibility of realizing complementary metal-oxide-semiconductor (CMOS) logic. To utilize the full potential of TMDs for the FET applications, charge doping is particularly important. Doping by vacancies or substitutional impurity atoms in the monolayer $MoS_2$ has been reported [10]. Doping by surface adatoms is a unique possibility for low dimensional materials. P-type charge transfer by the 7[th] column adatoms was reported in the chlorine (Cl) plasma reaction with graphene and graphene nanoribbons [11]. Moreover, doping via surface nitrogen-dioxide ($NO_2$) molecules has been shown to effectively dope monolayer $WSe_2$ p-FETs [9].

In this work, we examine the possible adatom doping of monolayer $MoS_2$ by means of density functional theory (DFT) calculations. We consider several atoms from the 1st column of the periodic table—lithium (Li), sodium (Na) and potassium (K)—and from the 7th column—fluorine (B), chlorine (Cl) and bromine (Br)—and metals—scandium (Sc), titanium (Ti), tantalum (Ta), palladium (Pd), platinum (Pt), silver (Ag), gold (Au), and molybdenum (Mo), itself. The electropositive alkaline elements from the 1st column are susceptible to donating electrons, while halogens from the 7th column are susceptible to accepting electrons. However, as we show here, these qualitative expectations are not necessarily substantiated by DFT calculations. The metal atoms we investigate have been experimentally examined as a metal contacts for TMDs [5,9]. In addition to possible functionality as a dopant, we investigate the most stable configuration of each adatom on monolayer $MoS_2$, the bonding type between adatom and the Mo and/or S atoms, and the amount of charge transfer from the adatom to monolayer $MoS_2$, all with the help of band structure, density of states (DOS), and charge density distributions.

Spin-polarized DFT calculations were performed using the projector-augmented wave method with a plane-wave basis set as implemented in the Vienna *ab initio* simulation package (VASP) [12]. The generalized gradient approximation (GGA) was applied for the exchange-correlation potential [13], which has been shown to reproduce the experimental band gap of monolayer TMDs well [14,15,16]. Semicore pseudopotentials were used for Mo ($4p^6 4d^5 5s^1$), Li ($1s^2 2s^1$), Na ($2p^6 3s^1$), K ($3s^2 3p^6 4s^1$), Sc ($3s^2 3p^6 3d^1 4s^2$), Ti ($3p^6 3d^2 4s^2$) and Ta ($5p^6 5d^3 6s^2$), and valence-only pseudopotential was used for S ($3s^2 3p^4$), F ($2s^2 2p^5$), Cl ($3s^2 3p^5$), Br ($4s^2 4p^5$), Pd ($4d^9 5s^1$), Pt ($5d^9 6s^1$), Ag ($4d^{10} 5s^1$) and Au($5d^{10} 6s^1$). We chose a kinetic energy cutoff of 400 eV. The *k*-mesh grid of 7×7×3 for the sampling of the 1st brillouin zone (BZ) of the super-cell was selected according to Monkhorst-Pack type meshes with the origin being at the Γ point for all calculations except the band structure calculation. In the band structure calculation, *k*-points along high symmetry directions (K−Γ−M) were used. For the electronic optimization, convergence was considered to be achieved when the change both in the total energy and in the eigenvalues between two successive self-consistent steps were smaller than $1\times10^{-5}$ eV. Corrections to the dipole moment and electrostatic potential were applied in the calculations.

It has been reported that the experimental lattice parameters well produce the band gap of monolayer $MX_2$ [14,15,16]. Therefore, we used the experimental in-plane lattice constant of *a* = 0.316 nm for the in-plane hexagonal primitive unit cell of $MoS_2$ monolayers absent any adatoms, as shown in Fig. 1(a) [17]. The adatom-monolayer $MoS_2$ system was modeled using one adatom in hexagonal super-cell of in-plane lattice constant 3*a* = 9.48 Å, as also shown in Fig. 1(a). The associated sheet doping density is quite large, $1.28\times10^{14}/cm^2$. However, we also considered both smaller (2*a* in plane lattice constant) and larger (4a in-plane lattice constants) super-cells for selected simulations and found little change in the energies of the relevant adatom-induced states beyond the 3*a* super-cell size. Note that the resulting BZ of the super-cell is correspondingly smaller and, in particular, the K point of primitive unit cell of the adatom-free system—where the band edges of $MoS_2$ are located—folds onto the Γ point of BZ of the super-cell. Except for retaining the fixed in-plane cell size, positions of all atoms were allowed to vary in all three dimensions as required to minimize the inter-atom forces. All atomic positions in the slabs were optimized according to a conjugate gradient minimization of the Hellman-Feynman forces until the magnitude of the force on each atom was 0.01 eV/Å or less. A vacuum region of ~20 Å was introduced between $MoS_2$ monolayers in the *z*-direction of the three-dimensional simulation unit cells to avoid layer-to-layer interactions. All calculations were performed at zero degrees Kelvin, and the highest occupied energy state was taken as the zero energy reference for the resulting band structure. For reference, the resulting calculated band gap of monolayer MoS$_2$ of 1.8 eV, as seen in Fig. 1(b), is close to values reported in previous studies [2,14, 15,16,18].

Calculations of adatom-monolayer MoS$_2$ system are carried out for the adatom located near, but allowed to relax in three dimensions about, three sites of high symmetry, as indicated in Fig. 1(a): top of Mo: *M*, top of S: *S*, and top of hexagon center: *H*. We calculate the binding energy from the equation

$$E_{\text{binding}} = E_{\text{MoS}_2} + E_{\text{adatom}} - E_{\text{MoS}_2+\text{adatom}},$$

where $E_{\text{binding}}$ is the binding energy of adatom on the monolayer MoS$_2$, $E_{\text{MoS}_2}$ is the energy of monolayer MoS$_2$ super-cell without the adatom, $E_{\text{adatom}}$ is the energy of isolated atom without monolayer MoS$_2$, and $E_{\text{MoS}_2+\text{adatom}}$ is the total energy of adatom-monolayer MoS$_2$ system. Of the three sites considered, the site with the largest binding energy (minimum total energy) is taken to be the most stable site. We then use that most stable structure to obtain the band structure, DOS and charge density. The charge transfer from the adatom to the MoS$_2$ is calculated as the difference between the number of valence electrons in the isolated adatom and the fractional number of electrons in the adatom in the adatom-monolayer MoS$_2$ super-cell by Bader's method [19], which partitions the charge density in a molecule to atoms according to the zero flux surface.

Our results are summarized in Table I where the most favored site, the bonding type, the charge transfer amount, and the type of doping if any is provided for each considered adatom to the MoS$_2$. The type of doping if any is determined by the resulting location of the highest occupied state: in the conduction band corresponds to n-type doping; in the valence band to p-type doping, and in the bandgap of MoS$_2$ to just being a trap/recombination center.

All of the considered adatoms from the 1st column, Li, Na and K, turn out to be effective n-type dopants in $MoS_2$ in these calculations. Each of these 1st column adatoms resides most stably at the *M* site, as illustrated in Fig. 2(a) for Li adatoms. The bonding of these adatoms also appears to be primarily ionic, as illustrated by the limited charge density in the region between the Li atoms and the M and S atoms in Fig 2(a), which is consistent with the previous experimental studies of 1st column K and Cs deposition on $MoS_2$ [20,21,22]. The Fermi level (The highest occupied state when within a continuous energy band) is pulled into the conduction band, and without any significant changes to the band gap or near-band edge band structure, producing mobile electrons. This behavior is illustrated in Fig. 3(a) where the band structure and atom projected DOS plots of monolayer $MoS_2$ with Li adatoms, specifically, are shown, along with the band structure and total DOS for the adatom-free system for reference.

In contrast, none of the considered 7th column adatoms are found to be effective dopants. Each of these 7st column adatoms resides most stably at the *S* site, as illustrated in Fig. 2(b) for F adatoms. The bonding of these 7th column adatoms also appears to be covalent, as illustrated by the substantial charge density between the F and S atoms in Fig 2(b). However, each F adatom also gains about 0.6 electron from the monolayer $MoS_2$ suggesting a relatively significant ionic bonding component as well for this particular adatom. These adatoms tend to produce two energy levels of what would otherwise be the band gap of $MoS_2$, a lower occupied state at zero Kelvin (donor like in this sense but far from the conduction band) and an upper empty state (acceptor like but far from the valence band), pinning the Fermi level between them, as illustrated for F in Fig 3(b). (Note that in these zero temperature calculations, the exact position of the Fermi level is indeterminate.) In this way, these states act like gold (Au) or copper (Cu) impurities in silicon (Si). With increasing atomic number of the adatom, these two states move somewhat downward within the gap. The energy gap between the lowest lying of these two energy levels and the valence band maximum is 0.469 eV, 0.322 eV and 0.287 eV for F, Cl, and Br adatoms, respectively.

Of these metal atoms considered, similarly, none appears to be clearly effective dopants in the simulated system. Of these metals, Sc, Ti, Ag reside most favorably on the *H* site and Au is most stable on the *S* site; all other reside most favorably on the *M* site. All exhibit covalent bonding. In addition, Sc, Ti and, to a lesser degree, Mo, Pd, Pt exhibit relatively significant ionic bonding components as well.

A similar approach was recently used to study the functionalization of monolayer $MoS_2$ through adatom adsorption in Ref [23]. We briefly discuss the similarities and differences between our study and Ref. [23] results. Among eight metal atoms considered in our work, Sc, Ti, Mo, and Pt were also investigated in Ref [23]. The most stable adatom site, namely *H* for Sc, and *M* for Mo and Pt are the same in both studies. While Ref. [23] suggests *M* as the most stable site for Ti, our results indicate *H* having a slightly higher binding energy. The locations of localized energy states induced by Sc, Mo and Pt adatoms agree fairly well with Ref [23]. Both calculations show a similar level of negative charge transfer from Sc and Ti to the monolayer $MoS_2$. However, contrary to our simulations predicting positive and negative charge transfer to monolayer $MoS_2$ from Mo and Pt, respectively, Ref [23] shows significant and some negative charge transfer from Mo and Pt to monolayer $MoS_2$, respectively. Our simulations also result in the larger binding energies for all four atoms. These discrepancies may arise from the different simulation conditions including use of a different in-plane lattice constant for monolayer $MoS_2$ and, perhaps, a different choice of pseudopotential. In Ref [23], a DFT-optimized in-plain lattice constant of 0.320 nm is reported for monolayer $MoS_2$, along with an associated band gap of 1.58 eV, while we fixed the in-plane lattice constant to the experimental value of 0.316 nm [17] in our calculations, which reproduces the measured band gap of 1.8~1.9 eV [2,3]. However, if we use the 0.320 nm lattice constant value from of [23], we also essentially recover the band gap of Ref. [23], finding a 1.589 eV band gap. Moreover, although the pseudopotential choice is not specified in Ref [23], in our test simulations using the 0.320 nm lattice constant and valence-only pseudopotentials which freeze the semicore electrons for all atoms, adatom and in the $MoS_2$ monolayer, we obtain similar binding energies and amounts of charge transfer as reported in [23].

We note that low (high) work function metals could be expected to transfer (extract) electrons to (from) the $MoS_2$ monolayer. These expectations, however, are only roughly borne out by the DFT simulations, as illustrated in Fig. 5. Moreover, none of this charge transfer actually results in free carriers in the nominal conduction or valence bands of the $MoS_2$ monolayer. However, one may speculate that, most promisingly, Sc and Pd could be brought closer to the band edges to serve as a donor or acceptor, respectively, in higher dielectric environment to the extent these states may be (2D) hydrogen-like.

In summary, we use density functional theory based electronic structure methods to study the effect of adatoms on monolayer $MoS_2$. Of the 1$^{st}$ column atoms considered for this purpose, Li, Na and K, all appear to be potentially effective n-type/donor dopants. However, none of the 7$^{th}$ column atoms considered, F, Cl, Br, none appear not to be effective dopants. Rather, they provide deep level traps much like Cu or Ag in Si. Among the various metal adatoms considered, Sc, Ti, Ta, Mo, Pd, Pt, Ag, Au, our calculations suggest no clear effective dopant under the considered simulation conditions. However, one may speculate that Sc and Pd might be able to serve as donors and acceptors respectively in a higher dielectric environment.

The authors acknowledge support from the Nanoelectronics Research Initiative supported Southwest Academy of Nanoelectronics (NRI-SWAN) center, and a grant from Intel. We thank the Texas Advanced Computing Center (TACC) for computational support.

**TABLE I**. Adatom type, adatom, most stable site, binding energy, bonding type—ionic, covalent, or covalent plus relatively significant ionic (covalent+)—, fractional charge transfer from the adatom to the MoS$_2$ monolayer, and whether and what type of significant mobile charge results in adatom-monolayer MoS$_2$ system.

| Adatom | | Site | Binding Energy [eV] | Bonding Type | Charge Transfer | Significant Mobile Charge |
|---|---|---|---|---|---|---|
| 1st | Li | H | 2.662 | | | |
| | | M | 2.817 | Ionic | -0.6700e | Electron |
| | | S | 2.189 | | | |
| | Na | H | 2.162 | | | |
| | | M | 2.168 | Ionic | -0.6449e | Electron |
| | | S | 1.850 | | | |
| | K | H | 2.361 | | | |
| | | M | 2.386 | Ionic | -0.8245e | Electron |
| | | S | 2.087 | | | |
| 7th | F | H | 1.197 | | | |
| | | M | 1.026 | | | |
| | | S | 1.671 | Covalent+ | 0.5721e | No |
| | Cl | H | 0.518 | | | |
| | | M | 0.410 | | | |
| | | S | 0.770 | Covalent | 0.2237e | No |
|       |     | H | 0.395 |           |          |     |
|       | Br  | M | 0.307 |           |          |     |
|       |     | S | 0.591 | Covalent  | 0.2172e  | No  |
|       |     | H | 3.449 | Covalent+ | -1.2807e | No  |
|       | Sc  | M | 3.292 |           |          |     |
|       |     | S | 2.292 |           |          |     |
|       |     | H | 3.712 | Covalent+ | -1.1788e | No  |
|       | Ti  | M | 3.686 |           |          |     |
|       |     | S | 1.900 |           |          |     |
|       |     | H | 3.338 |           |          |     |
|       | Ta  | M | 3.696 | Covalent  | 0.0884e  | No  |
|       |     | S | 2.563 |           |          |     |
|       |     | H | 2.181 |           |          |     |
|       | Mo  | M | 2.325 | Covalent+ | 0.5077e  | No  |
| Metal |     | S | 1.764 |           |          |     |
|       |     | H | 3.000 |           |          |     |
|       | Pd  | M | 3.344 | Covalent+ | 0.3751e  | No  |
|       |     | S | 2.819 |           |          |     |
|       |     | H | 3.411 |           |          |     |
|       | Pt  | M | 3.975 | Covalent+ | -0.5800e | No  |
|       |     | S | 3.557 |           |          |     |
|       |     | H | 1.731 | Covalent  | 0.1558e  | No  |
|       | Ag  | M | 1.685 |           |          |     |
|       |     | S | 1.578 |           |          |     |
|       |     | H | 1.847 |           |          |     |
|       | Au  | M | 1.761 |           |          |     |
|       |     | S | 1.904 | Covalent  | 0.1113e  | No  |

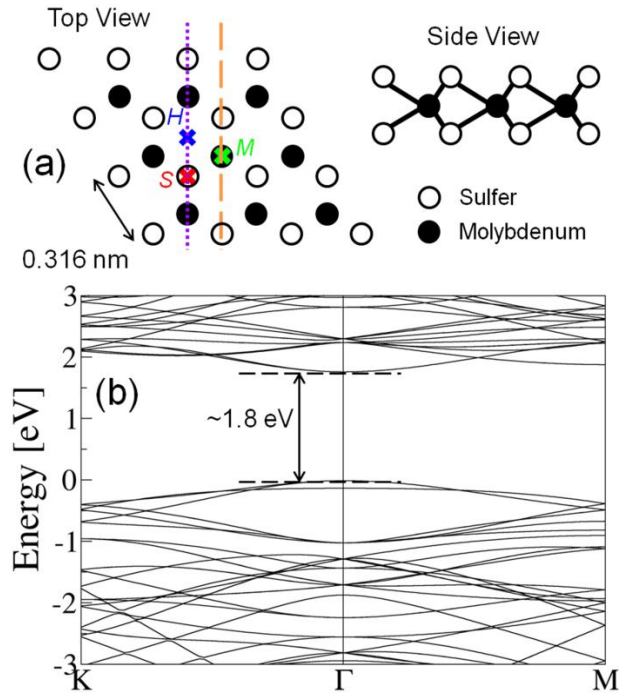

**FIG. 1**. (a) Top and side views of unit super-cell of monolayer MoS$_2$ and (b) corresponding band structure along high symmetry directions of the hexagonal BZ. Three adatom sites are considered within this super-cell as indicated: top of Mo = *M*, top of S = *S*, and top of hexagon center = *H*. The dotted and dashed lines indicate the location of charge density slices shown in Fig. 2.

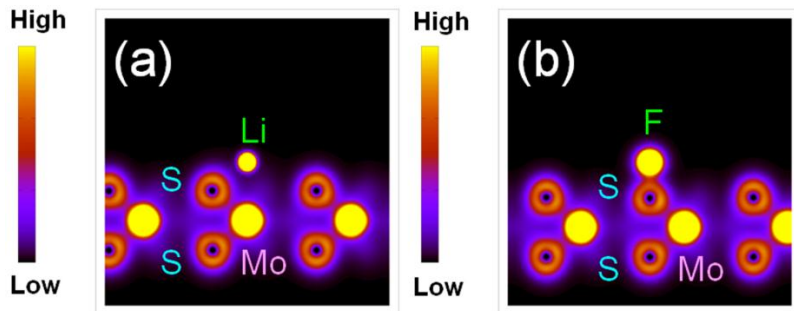

**FIG. 2**. Charge densities on slices perpendicular to the surface of monolayer MoS$_2$, located as shown in Fig. 1, for (a) Li and (b) F adatoms.

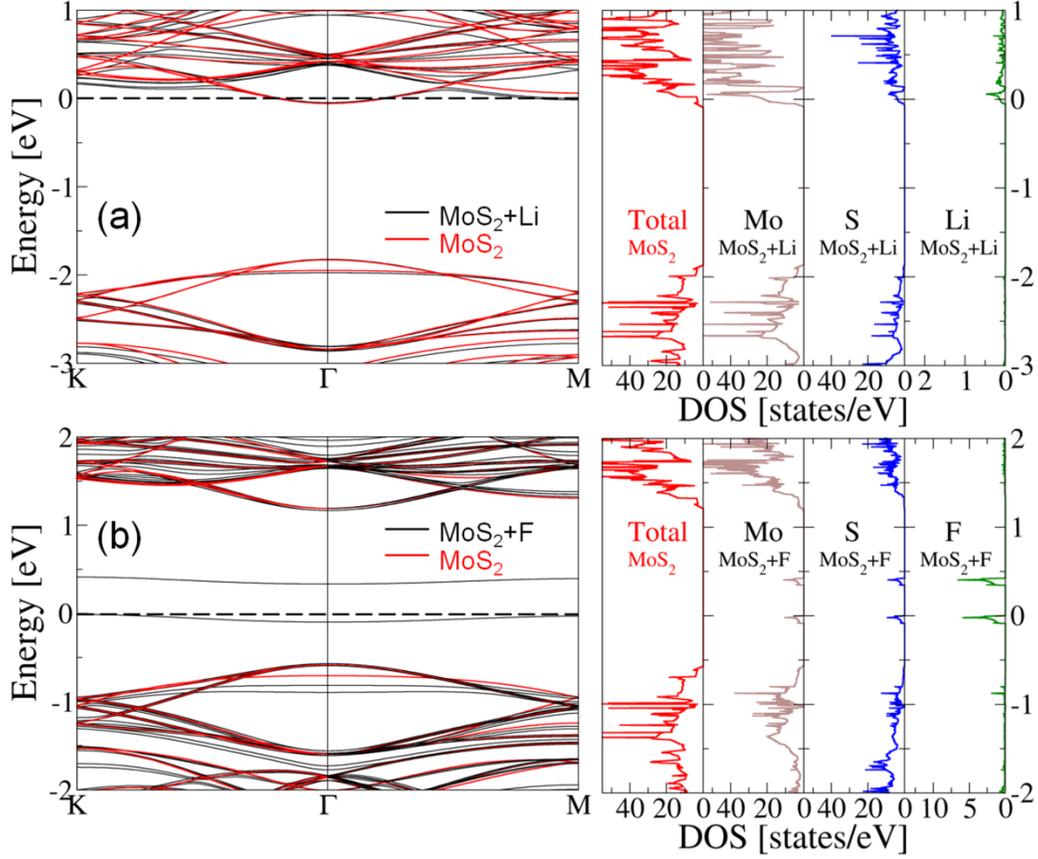

**FIG. 3**. Band structures (black lines, left-hand-side (LHS)) along high symmetry directions of the hexagonal BZ and atom-projected DOS (right-hand-side (RHS)) of monolayer $MoS_2$ for (a) Li and (b) F adatoms. Also provided for reference is band structure (red lines, LHS) and total DOS for the adatom-free monolayer $MoS_2$ super-cell.

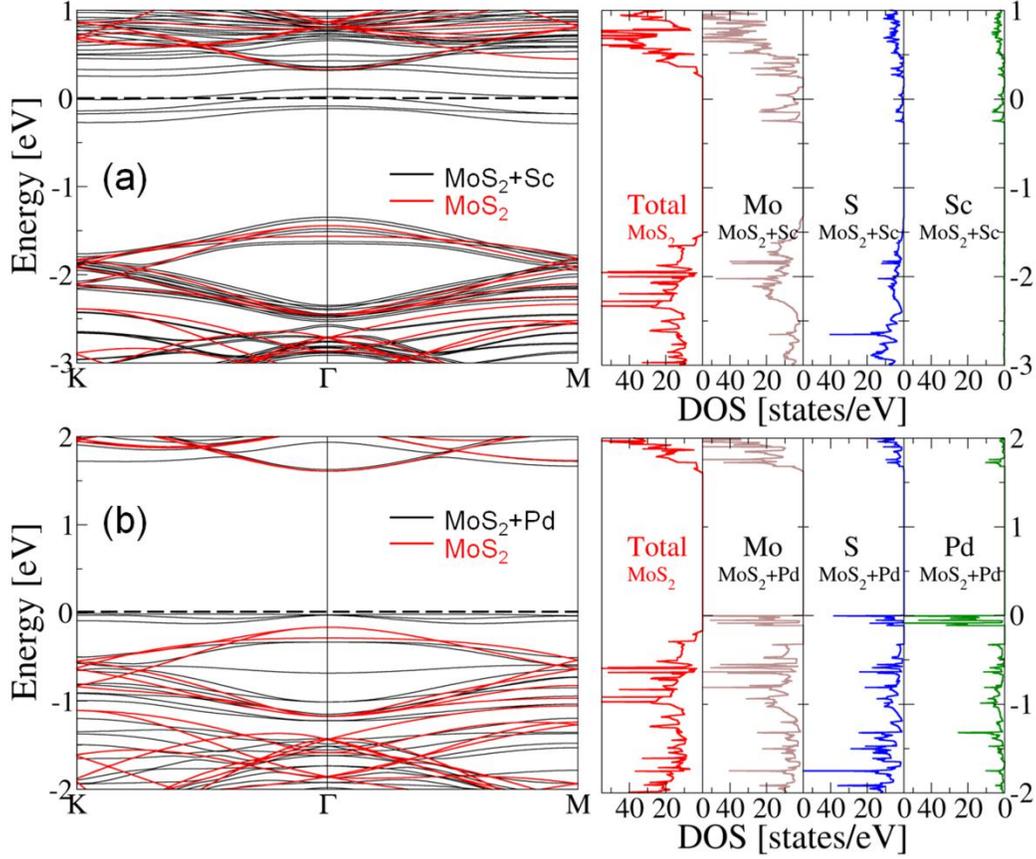

**FIG. 4**. Band structures (LHS) along high symmetry directions of the hexagonal BZ and atom projected densities of states (RHS) of monolayer $MoS_2$ (a) with Sc and (b) with Pd. Also provided for reference is band structure (red lines, LHS) and total DOS for the adatom-free monolayer $MoS_2$ super-cell. Note that for both Sc and Pd, the states created within the $MoS_2$ band gap by the adatoms are donor-like in the sense that they are occupied in the low temperature limit.

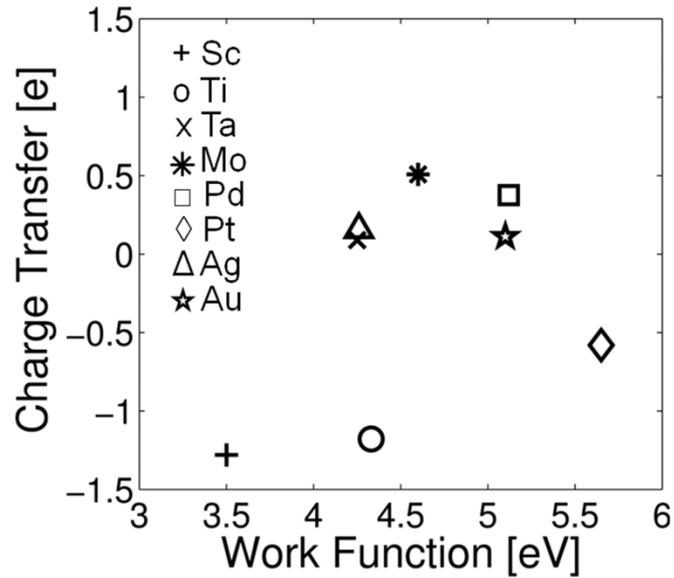

**FIG. 5**. Charge transfer from adatom to MoS$_2$ monolayer vs. work function for the metal adatoms considered here.